\documentclass[10pt]{article}

\usepackage{graphicx}
\def \to {\rightarrow}
\def\xslash#1{{\rlap{$#1$}/}}

\def\bfsig{\mbox{\boldmath$\sigma$}}

\newsavebox{\hflrar}
\sbox{\hflrar}{\makebox[0pt][l]
{${\scriptstyle \leftharpoonup}$}{${\scriptstyle \rightharpoonup}$}}
\def \hd {\raisebox{1.8ex}{\usebox{\hflrar}} \hspace*{-8pt} D}
\def \hbd {\raisebox{1.8ex}{\usebox{\hflrar}} \hspace*{-8pt} {\bf D}}

\begin{document}
\begin{center}
{\Large Corrections For Two Photon Decays of $\chi_{c0}$ and $\chi_{c2}$ and Color Octet Contributions}
\vskip 10mm
J. P. Ma and Q. Wang   \\
{\small {\it Institute of Theoretical Physics , Academia
Sinica, Beijing 100080, China }} \\
\end{center}

\vskip 1 cm


\begin{abstract}
Using the fact that the c-quark inside a charmonium moves with a small
velocity $v$ in the charmonium rest-frame, one can employ an expansion
in $v$ to study decays of charmonia and results at the leading order
for $\chi_{c0,2}\to\gamma\gamma$ exist in the literature. We study corrections
at the next-to-leading order in the framework of nonrelativistic QCD(NRQCD) factorization.
The study presented here is different than previous approaches where $\chi_{c0,2}$
is taken as a bound-state of a $c\bar c$ pair and a nonrelativistic wave-function
is used for the pair.
We find that the corrections are consist not only of relativistic corrections,
but also of corrections from Fock state components of $\chi_{c0,2}$ in which the
$c\bar c$ pair is in a color-octet state. For $\chi_{c2}$ there is also a contribution
from a Fock state component in which the pair is in a F-wave state. We determine
the factorization formula for decay widths in the form of NRQCD matrix elements representing
nonperturbative effects related to $\chi_{c0,2}$, and calculate the perturbative
coefficients at tree-level. Because the NRQCD matrix elements are unknown, a detailed
prediction for the decay $\chi_{c0,2}\to\gamma\gamma$  can not be made, but the effect
of these corrections can be determined at certain level. Estimations show that
the effect is significant and can not be neglected.
\vskip 5mm

\end{abstract}

\vfill\eject\pagestyle{plain}\setcounter{page}{1}

\par\noindent
{\bf 1. Introduction}
\par\vskip20pt
\par
Because of their simple final states, decays of P-wave charmonium into two photons can be
good channels to determine the value of $\alpha_s$ and for study of properties
of charmonium. In recent years various experiment groups have published
their results on $\Gamma_{\gamma \gamma}(\chi_{c0})$ and $\Gamma_{\gamma
\gamma}(\chi_{c2})$~\cite{835,CLEO2001,CLEO1994,L3,OPAL}.
Some major and newest results are listed in Table 1, where errors from different sources
are combined.
From these data it is easy to find that although there is a tendency of
consistency in latest results of CLEO and E835, values of different groups
have a large discrepancy from each other. On the theoretical side, because charm quark
can be taken as heavy quark and it moves with a small velocity $v$ inside
a charmonium in its rest-frame, one may describe the $c$- or $\bar c$ quark
inside charmonia with a nonrelativistic wave-functions by taking charmonia as a bound-state
of a $c\bar c$ pair. In most previous calculations for decays of charmonium
such a nonrelativistic wave-functions is employed for
$\chi_{c0,2}\to \gamma\gamma$\cite{Barb,Berg,Barnes,Munz,Zhao}.
By expanding the small velocity $v$, the decay
width  at the leading order of $v$
can be  expressed with the derivatives of the wave functions at origin, where
the leading order is at $v^2$. The corrections from the next-to-leading order
of $v$ are also studied with the nonrelativistic wave-functions\cite{Berg,Barnes,Munz,Zhao},
in which the corrections are only relativistic corrections. This also implies
that at the next-to-leading order the charmonia $\chi_{c0,2}$ are still taken as a
bound-state of a $c\bar c$ pair and the pair has the same quantum number of $\chi_{c0,2}$.
\par
\begin{table}
\caption{List of major experimental results}
\begin{center}
\begin{tabular}{|l|l|l|}
\hline
     &  $\Gamma_{\gamma \gamma}(\chi_{c0}) (keV)$  & $\Gamma_{\gamma \gamma}
(\chi_{c2}) (keV)$  \\
\hline
CLEO(1994) &                   & 1.08 $\pm$ 0.28 \\
CLEO(2001) & 3.76 $\pm$ 1.07   & 0.53 $\pm$ 0.16  \\
E835(2000) & 1.61 $\pm$ 0.75   & 0.270 $\pm$ 0.042 \\
L3(1999)   &                   &  1.02 $\pm$ 0.25 \\
OPAL(1998) &                   & 1.76 $\pm$  0.37  \\
\hline
\end{tabular}
\end{center}
\end{table}

\par

From point of view of a quantum field theory like QCD, a hadron state is a bound state
of quarks and gluons and has many components composed with different number of quarks
and of gluons. For charmonia it can be expanded in $v$ and for
$\chi_{c0,2}$ it looks like:
\begin{equation}
|\chi_{c0,2} \rangle = | (c \bar c )_1 \rangle + O(v) | (c \bar c )_8 g \rangle + higher\ orders,
\end{equation}
where the index 1 or 8 denotes color singlet or color octet respectively.
From this expansion the state $| (c \bar c )_1 \rangle$ is a dominant component,
while other components have a suppressed probability to be observed in
$\chi_{c0,2}$, the suppression parameter is $v$. Hence, in a systematic study
of higher-order corrections one should also include contributions from
these suppressed components, i.e., a systematic expansion in $v$ is needed.
Recently, a novel approach based on NRQCD for productions and decays of quarkonia
is proposed\cite{Bodwin}. In this approach a decay width can be systematically
expanded in the velocity $v$. For example, the decay width of $\chi_{c0}$
can be written as:
\begin{equation}
\Gamma(\chi_{c0}\to\gamma\gamma) = \sum_n c_n \langle \chi_{c0} \vert
   \hat O_n \vert\chi_{c0}\rangle,
\end{equation}
where $\hat O_n$'s are operators defined in NRQCD and $c_n$ can be calculated with
perturbative theory. There is a rule of power counting of $v$ for the operators
$\hat O_n$\cite{PC}.
We will follow this approach to study
the next-to-leading order corrections for the decay $\chi_{c0,2}\to \gamma\gamma$.
We will determine the form of relevant operators in NRQCD and calculate the corresponding
coefficients. It should be noted that the role of the suppressed component
$| (c \bar c )_8 g \rangle$ has been studied before in inclusive processes. It has been
shown\cite{Bodwin2} that only by including the contribution from $| (c \bar c )_8 g \rangle$
the inclusive decay of $\chi_{cJ}$ can be consistently predicted with QCD. It has been
also shown that such components with a color octet $c\bar c$ pair of $\psi$ are
very important for explaining the $\psi$ surplus at the Tevatron\cite{BF}.
In this work we will show the effect of such components in exclusive processes. The effect
of $| (c \bar c )_8 g \rangle$ has been studied in $\chi_{cJ}\to\pi\pi$\cite{PIPI}, in which
a light-cone wave-function for the component was introduced and the effect was not
parameterized with matrix elements of NRQCD. 
\par
At the next-to-leading order of $v$, the corrections, as we will see, consist of
relativistic corrections and also corrections from the state $| (c \bar c )_8 g \rangle$.
Beside these, for $\chi_{c2}$ there is also a correction from the state
$| (c \bar c )_1 \rangle$ where the $c\bar c$ pair is in a state with the orbit angular
momentum $l=3$. These corrections take a factorized form as products of pertubative coefficients and
matrix elements,
in which the matrix elements are defined in NRQCD and they parameterize nonperturbative effects.
We will determine the form of these matrix element and calculate these coefficients at leading order
of $\alpha_s$. Our work is organized as the following: In Sect. 2 we calculate
these coefficients and give our main results. In Sect. 3 we discuss the effect of these
corrections and Sect.4 is our summary.
\par\vskip20pt
\par\noindent
{\bf 2. The correction at the next-to leading order}
\par\vskip20pt
We consider $\chi_{c0,2}$ to two photon decay in its rest frame:
\begin{equation}
\chi_{c0,2} (P) \longrightarrow \gamma (k_1)\ \ +\ \ \gamma (k_2)
\end{equation}
where P, $k_1$, $k_2$ denote momentum of $\chi_{c0,2}$ and two photons.
We decompose the decay width as
\begin{equation}
\Gamma =\Gamma_1+\Gamma_8,
\end{equation}
where $\Gamma_{1,8}$ denotes the contribution from a color-singlet- and color-octet
$c\bar c$ pair, respectively. The color-singlet contribution
can be written as:
\begin{eqnarray}
 \Gamma_1=&16& {Q_c}^4 e^4 \int d \Gamma \int \frac{d^4 p_1}{(2\pi)^4} \frac{d^4
{p_1}^\prime}{(2\pi)^4} \int d^4 x e^{i 2(k_1+k_2)x}
\int d^4 y e^{-i p_1(x-y)}\langle 0| \bar {c_i} (x) c_j(y) |\chi_{c0,2} \rangle \nonumber \\
&\times & \int d^4 y^\prime e^{-i {p_1}^\prime (y^\prime +x)} \langle \chi_{c0,2} |
\bar {c_{j^\prime}} (y^\prime) c_{i^\prime} (-x) |0 \rangle \times
H_{ij,i^\prime j^\prime} \\
&& d \Gamma = \frac{d^4 k_1}{(2\pi)^4} \frac{d^4 k_2}{(2\pi)^4} (2\pi)^2
\delta_+ ({k_1}^2) \delta_+ ({k_2}^2) \\
&& H_{ij,i^\prime j^\prime} = {[\gamma^\mu \frac{1}{\xslash {p_1}
-\xslash {k_2} -m_c} \gamma^\nu + \gamma^\nu \frac{1}{\xslash {p_1} -
\xslash {k_1} -m_c} \gamma^\mu]}_{ij} \times
{[\gamma_\nu \frac{1}{\xslash {p_1}^\prime -\xslash {k_2} -m_c} \gamma_\mu ]}
_{j^\prime i^\prime},
\end{eqnarray}
where $c(x)$ stands for the Dirac field of $c$-quark and $i,j,i^\prime$ and
$j^\prime$ stand for color- and spin indices. This
contribution can be represented by the diagrams in Fig.1. The matrix element
with Dirac fields of $c$-quark represents the nonperturbative effect related
to the initial hadron. The quantity $H$ is calculated with perturbative theory.
An expansion in $v$ can be performed by expanding
the Dirac fields with fields of NRQCD. This can be done by using
Foldy-Wouthuysen transformation\cite{fw}:

\begin{figure}
\includegraphics[width=6in]{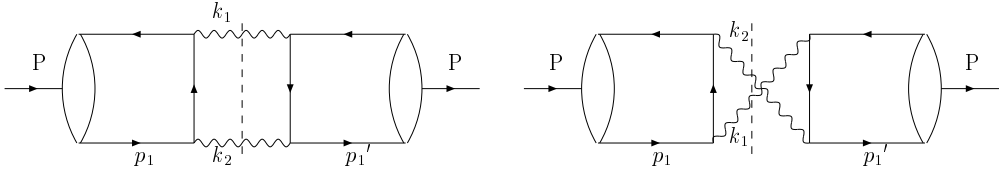}
\caption{Diagrams of contribution for decay width from $|(c \bar c)_1 \rangle$ state}
\label{fig:one}
\end{figure}

\begin{eqnarray}
 c(x) &=& e^{-i m_c t} [ 1-\frac{i\gamma_j D_j}{2m_c} +\frac{1}{8{m_c}^2}
(-i \gamma_j D_j)^2 + \frac{-1}{4{m_c}^2} \gamma_j [D_0, D_j] + \frac{3}{16{m_c}^3}
(-i \gamma_j D_j)^3 ] \psi(x)  \nonumber \\
 &+& e^{i m_c t} [ 1- \frac{i\gamma_j D_j}{2m_c} +\frac{1}{8{m_c}^2} (-i \gamma_j D_j)^2
+\frac{1}{4{m_c}^2} \gamma_j [D_0,D_j] +\frac{3}{16{m_c}^3} (-i \gamma_j D_j)^3 ]\chi(x)
 + O(v^4) \nonumber \\
\ \
\end{eqnarray}
where $\psi(x)$ and $\chi(x)$ are NRQCD fields and $\psi(\chi^\dagger)$
annihilates a $c(\bar c)$-quark respectively. $m_c$ is the pole mass of $c$-quark.
Using this transformation the matrix elements are then of the NRQCD
fields $\chi$, $\psi$ and their derivatives. Operators ${\bf D}$ and $\partial _i
$ are of $v^1$ order and $D_0$ and $\partial _0$ are of $v^2$ order in NRQCD
according the power counting rule\cite{Bodwin}. The space-time dependence
of matrix element are controlled with different scales in the different directions.
In the time direction it is $m_c v^2$, while in the spacial direction it is
$m_c v$. The expansion in $v$ is then completed by doing
Taylor expansion of  fields $\chi$ and $\psi$ around origin. Using the power
counting rule in \cite{Bodwin} we collect all local matrix elements up to
order of $v^4$, which are allowed by rotation-, charge conjugation- and party symmetries
for $\chi_{c0,2}$,
and calculate the corresponding contributions. Doing the calculation in this
way is similar to the matching method by taking a free $c\bar c$ pair, as proposed
in \cite{Bodwin}, the difference is that we only collect the contributions
from operators with correct quantum numbers in the problem considered here.
The calculation is rather tedious but straightforward. We obtain:
\begin{eqnarray}
 \Gamma_1 (\chi_{c0} \longrightarrow \gamma \gamma) &=& \frac{6\pi {Q_c}^4
{\alpha}^2}{{m_c}^4} \langle \chi_{c0} | {\cal O}_{EM} (^3 P_0) | \chi_{c0} \rangle
-\frac{14\pi {Q_c}^4 \alpha^2}{{m_c}^6} \langle \chi_{c0} | {\cal G}_{EM} (^3 P_0)
| \chi_{c0} \rangle + {\cal O}(v^6) \nonumber \\
&\ &\  \\
 {\cal O}_{EM} (^3 P_0) &=& \frac{1}{3} \psi^\dag ( \frac{-i}{2} \hbd
\cdot {\bfsig} ) \chi |0 \rangle \langle 0| \chi^\dag (\frac{-i}{2}
\hbd \cdot {\bfsig} ) \psi \\
 {\cal G}_{EM} (^3 P_0) &=& \frac{1}{2} [\frac{1}{3} \psi^\dag (\frac{-i}{2}
\hbd)^2 (\frac{-i}{2} \hbd\cdot {\bfsig}) \chi |0
\rangle \langle 0| \chi^\dag (\frac{-i}{2} \hbd  \cdot {\bfsig})
\psi + h.c.],
\end{eqnarray}
where $\sigma_i(i=1,2,3)$ is the Pauli matrix. With the above results we find the
correction to the color-singlet contribution is only the relativistic correction.
Similarly, the decay width of $\chi_{c2}$ is:
\begin{eqnarray}
 \Gamma_1(\chi_{c2} \longrightarrow \gamma \gamma) &=& \frac{8\pi {Q_c}^4
\alpha^2}{5{m_c}^4} \langle \chi_{c2} | {\cal O}_{EM} (^3 P_2) | \chi_{c2} \rangle
- \frac{172\pi {Q_c}^4 \alpha^2}{105{m_c}^6} \langle \chi_{c2} | {\cal G}_{EM}
(^3 P_2) |\chi_{c2} \rangle \nonumber \\
&-& \frac{284\pi {Q_c}^4 \alpha^2}{105{m_c}^6} \langle \chi_{c2} | {{\cal G}^\prime
_{EM}} (^3P_2) | \chi_{c2} \rangle +{\cal O}(v^6) \\
 {\cal O}_{EM} (^3 P_2) &=& \psi^\dag (\frac{-i}{2} \hd
^{(i} \sigma ^{j)} ) \chi |0 \rangle \langle 0| \chi^\dag (\frac{-i}{2}
\hd{^{(i} \sigma^{j)}} ) \psi \\
 {\cal G}_{EM} (^3 P_2) &=& \frac{1}{2} [ \psi^\dag (\frac{-i}{2} \hbd)
^2 (\frac{-i}{2} \hd ^{(i} \sigma ^{j)}) \chi |0 \rangle \langle
0| \chi^\dag (\frac{-i}{2} \hd ^{(i} \sigma ^{j)}) \psi + h.c.] \\
{{\cal G}^\prime _{EM}} (^3P_2) &=& \frac{1}{2} [\psi^\dag (\frac{-i}{2} \hd )
^{(i} (\frac{-i}{2} \hd)^{j)} (\frac{-i}{2} \hbd\cdot {\bfsig}) \chi |0
\rangle \langle 0| \chi^\dag (\frac{-i}{2} \hd ^{(i} {\sigma}^{j)}) \psi +
h.c. ],
\end{eqnarray}
where the notation $(ij)$ means $T^{(ij)}=(T^{ij}+T^{ji})/2-\delta_{ij}T^{kk}/3$,
i.e., the tensor is symmetric and trace-less. Unlike the case with
$\chi_{c0}$ we find not only a relativistic correction but also
a contribution from a mixing with a $F$-wave $c\bar c$ pair. This is allowed
by symmetries, the $c\bar c$ pair is in a spin triplet state, a spin triplet
state with a orbit angular  momentum $l=3$ can have a total angular momentum
$J=2,3,4$. The state with $J=2$ is just the $\chi_{c2}$ state.
\par
Now we consider color octet contributions from $|(c \bar c)_8 g \rangle$. At
the order we consider the contributions can be represented by diagrams in
Fig.2. The contributions before the expansion in $v$ can be written as:
\begin{eqnarray}
\Delta \Gamma_{8a}&=& {Q_c}^4 e^4 \int d\Gamma \int d^4 y \int d^4 x
\frac{d^4 k}{(2\pi)^4} e^{i k (y-x)+i(k_1 +k_2)x} \langle 0|
\bar {c_i}(x) c_j(y) |\chi_{c0,2} \rangle \nonumber \\
&\times &\int d^4 y^\prime d^4 x^\prime \int \frac{d^4 k^\prime}
{(2\pi)^4} \frac{d^4 q}{(2\pi)^4} e^{ik^\prime y^\prime +i q x^\prime}
\langle \chi_{c0,2}| \bar {c_{i^\prime}}(0) g_s G^\mu (x^\prime) c_{j^\prime} (y^\prime)
|0 \rangle \times {M^a}_{ij,i^\prime j^\prime} \\
\Delta \Gamma_{8b}&=& {Q_c}^4 e^4 \int d\Gamma \int d^4 y \int d^4 x
\frac{d^4 k}{(2\pi)^4} e^{i k (y-x)+i(k_1 +k_2)x} \langle 0|
\bar {c_i}(x) c_j(y) |\chi_{c0,2} \rangle \nonumber \\
&\times & \int d^4 y^\prime d^4 x^\prime \int \frac{d^4 k^\prime}
{(2\pi)^4} \frac{d^4 q}{(2\pi)^4} e^{-ik^\prime y^\prime -i q x^\prime}
\langle \chi_{c0,2}| \bar {c_{i^\prime}}(y^\prime) g_s G^\mu (x^\prime) c_{j^\prime}(0)
|0 \rangle \times {M^b}_{ij,i^\prime j^\prime} \\
\Delta \Gamma_{8c}&=& {Q_c}^4 e^4 \int d\Gamma \int d^4 y \int d^4 x
\frac{d^4 k}{(2\pi)^4} e^{i k (y-x)+i(k_1 +k_2)x} \langle 0|
\bar {c_i}(x) c_j(y) |\chi_{c0,2}\rangle \nonumber \\
&\times &\int d^4 y^\prime d^4 x^\prime \int \frac{d^4 k^\prime}
{(2\pi)^4} \frac{d^4 q}{(2\pi)^4} e^{-ik^\prime y^\prime -i q x^\prime}
\langle \chi_{c0,2}| \bar {c_{i^\prime}}(y^\prime) g_s G^\mu (x^\prime) c_{j^\prime}(0)
|0 \rangle \times {M^c}_{ij,i^\prime j^\prime}
\end{eqnarray}
with
\begin{eqnarray}
{M^a}_{ij,i^\prime j^\prime}=&& {[\gamma^\alpha \frac{1}{\xslash k -
\xslash {k_2} -m_c + i \epsilon} \gamma^\beta + \gamma^\beta \frac{1}
{\xslash k - \xslash {k_1} -m_c +i\epsilon} \gamma^\alpha]}_{ij}
\nonumber\\
&& \times
{[\gamma_\beta \frac{1}{\xslash {k^\prime} + \xslash q +\xslash {k_1} -m_c
+i\epsilon} \gamma_\alpha \frac{1}{\xslash{k^\prime} +\xslash q -m_c +i\epsilon}
\gamma_\mu ]}_{i^\prime j^\prime} \nonumber \\
&&\ \   \\
{M^b}_{ij,i^\prime j^\prime}=&& {[\gamma^\alpha \frac{1}{\xslash k -
\xslash {k_2} -m_c + i \epsilon} \gamma^\beta + \gamma^\beta \frac{1}
{\xslash k - \xslash {k_1} -m_c +i\epsilon} \gamma^\alpha]}_{ij}
\nonumber\\
&&\times
{[\gamma_\mu \frac{1}{\xslash {k^\prime} + \xslash q -m_c +i\epsilon}
 \gamma_\beta \frac{1}{\xslash{k^\prime} +\xslash q -\xslash{k_2} -m_c +i\epsilon}
\gamma_\alpha ]}_{i^\prime j^\prime} \nonumber \\
&&\ \  \\
{M^c}_{ij,i^\prime j^\prime}=&& {[\gamma^\alpha \frac{1}{\xslash k -
\xslash {k_2} -m_c + i \epsilon} \gamma^\beta + \gamma^\beta \frac{1}
{\xslash k - \xslash {k_1} -m_c +i\epsilon} \gamma^\alpha]}_{ij}
\nonumber\\
&& \times
{[\gamma_\beta \frac{1}{\xslash {k^\prime} - \xslash{k_2}  -m_c +i\epsilon}
 \gamma_\mu \frac{1}{\xslash{k^\prime} +\xslash q -\xslash{k_2} -m_c +i\epsilon}
\gamma_\alpha ]}_{i^\prime j^\prime} \nonumber \\
&&\ \
\end{eqnarray}

\begin{figure}
\includegraphics[width=5in]{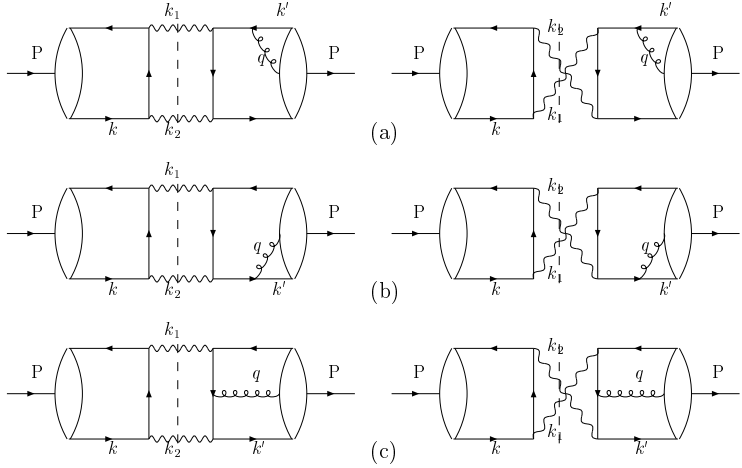}
\caption{Diagrams contribute to decay width from $|(c \bar c)_8 g \rangle$
state, the diagrams for conjugated contributions are not shown. }
\label{fig:two}
\end{figure}

Similarly we need to perform the expansion in $v$ for matrix element
like $\langle \chi_{c0}| \bar c_i(x) g_s G^\mu (y) c_j(0)|0 \rangle$.
Again with the power counting rule, one can find that the leading term
for the matrix element is related to the matrix element of the operator
$ \psi^\dagger g_s{\bf E}\cdot\bfsig \chi$. This matrix element is at order
of $v^3$ and will lead to a contribution to $\Gamma_8$ at order of $v^4$.
To calculate the contribution we
take the gauge: $G^0 =0$, in this gauge the electric chromofield is:
\begin{equation}
{\bf E}  = \frac{\partial}{\partial t} {\bf G}.
\end{equation}
Performing the expansion for matrix elements in Eq.(19), results can be
obtained in a straightforward way. We obtain the contribution at $v^4$:
\begin{eqnarray}
\Gamma_8 (\chi_{c0} \longrightarrow \gamma \gamma ) &=& \frac{-4\pi \alpha^2
{Q_c}^4} {{m_c}^5} \langle \chi_{c0} | {\cal F}_{EM} (^3 P_0 ) |\chi_{c0}
\rangle + {\cal O}(v^6) \\
 {\cal F}_{EM} (^3 P_0) &=&\frac{1}{6} [ \psi^\dag \bfsig \cdot
g_s {\bf E} \chi |0 \rangle \langle 0| \chi^\dag \bfsig \cdot
{\bf D} \psi + h.c. ] \\
\nonumber \\
 \Gamma_8 (\chi_{c2} \longrightarrow \gamma \gamma )&=& \frac{-2\pi {Q_c}^4
\alpha^2}{{m_c}^5} \langle \chi_{c2} | {\cal F}_{EM} (^3 P_2) | \chi_{c2} \rangle
+ {\cal O}(v^6)\\
 {\cal F}_{EM}(^3 P_2) &=& \frac{1}{2} [ \psi^\dag {\sigma}^{(i}
g_s {E}^{j)} \chi |0 \rangle \langle 0| \chi^\dag {\sigma }
^{(i} \hd^{j)} \psi + h.c. ].
\end{eqnarray}
\par
Eq.(9), Eq.(12), Eq.(23) and Eq.(25) are our main results. From these results we can see
that the corrections consist not only of relativistic correction which is represented
by the matrix elements $\langle \chi_{c0}\vert{\cal G}_{EM}(^3P_0)\vert\chi_{c0}\rangle$
and $\langle \chi_{c2}\vert{\cal G}_{EM}(^3P_2)\vert\chi_{c2}\rangle$, but also of contributions
from the color octet component. For $\chi_{c2}$ there is also a contribution from the
component in which the $c\bar c$ pair is in color single state with the orbit angular
momentum $l=3$.
\par\vskip20pt
\noindent
{\bf 3. Numerical impact of the correction}
\par\vskip20pt
To make theoretical predictions for decay widthes one needs to know all matrix elements
in the last section. Unfortunately, there is no information available for
the matrix elements appearing in the correction. There are estimations for
the matrix element at the leading order of $v$, but for consistency we need
an estimation for them at the accuracy of ${\cal O}(v^4)$. All this prevents us
from detailed predictions. However, numerical impact of the correction can be studied
by noting that there are several relations among these operators. Because of spin
symmetry of NRQCD we can have:
\begin{eqnarray}
\langle \chi_{c2} | {\cal O}_{EM}(^3P_2) |\chi_{c2} \rangle &=& \langle
\chi_{c0} | {\cal O}_{EM}(^3P_0) |\chi_{c0} \rangle ( 1+{\cal O}(v^2)),
\nonumber\\
\langle \chi_{c2} | {\cal G}_{EM}(^3P_2) |\chi_{c2} \rangle &=& \langle
\chi_{c0} | {\cal G}_{EM}(^3P_0) |\chi_{c0} \rangle ( 1+{\cal O}(v^2)),
\nonumber\\
\langle \chi_{c2} | {\cal F}_{EM}(^3P_2) |\chi_{c2} \rangle &=& \langle
\chi_{c0} | {\cal F}_{EM}(^3P_0) |\chi_{c0} \rangle ( 1+{\cal O}(v^2)).
\end{eqnarray}
As discussed before, the first relation is not useful here unless we know
the correction at ${\cal O}(v^2)$.
Beside these relations one can also
use equation of motion to obtain another type of relations. Generally there are
subtleties for using equation of motion for operators. However, it is shown
that one can still use equation of motion for operators which are sandwiched
between physical states\cite{Politzer}. For operators relevant to $J/\Psi$
one has used this to derive this type of relations\cite{Gremm}.
\par
The equation of motion of NRQCD reads:
\begin{equation}
(iD_t + \frac{{\bf D}^2}{2 m} )\psi =0, \hspace{1cm} (iD_t - \frac{{\bf D}^2}
{2 m} ) \chi =0.
\end{equation}
We can use these equations to replace the operator ${\bf D}^2$ with
$D_t$. For doing this we note that the matrix element appearing
in Eq.(10) can be written as:
\begin{equation}
\langle 0\vert \chi^\dag (\frac{-i}{2}\hbd\cdot\bfsig) (\frac{-i}{2}\hbd)^2 \psi \
\vert \chi_{c0} \rangle = \frac{i}{2}\langle 0 \vert
\chi^\dag {\bf D \cdot {\bf \sigma} D}^2 \psi
+({\bf D}^2 \chi)^\dag {\bf D \cdot {\bf \sigma}} \psi |\chi_{c0}\rangle
 \cdot \{1+{\cal O}(v^2)\},
\end{equation}
where the corrections terms are due to interchange of the gauge covariant
derivative ${\bf D}$, and they are at order of ${\cal O}(v^2)$ and can be neglected
here. Using the equation of motion we obtain:
\begin{eqnarray}
\langle 0| \chi^\dag {\bf D \cdot {\bf \sigma} D}^2 \psi
+({\bf D}^2 \chi)^\dag {\bf D \cdot {\bf \sigma}} \psi |\chi_{c0}\rangle
 &=& -2m_c i[\langle 0| \chi^\dag {\bf D \cdot {\bf \sigma}} D_t \psi
+ (D_t \chi)^\dag {\bf D \cdot {\bf \sigma}} \psi |\chi_{c0}
\rangle ] \nonumber \\
    &=&-2m_c E_{\chi_{c0}} \langle 0|\chi^\dag {\bf D \cdot {\bf \sigma}} \psi |
\chi_{c0} \rangle -2m_c \langle 0| \chi^\dag g{\bf E \cdot {\bf \sigma}} \psi |
\chi_{c0} \rangle,
\end{eqnarray}
where $E_{\chi_{c0}}$ is the binding energy. Doing the same for $\chi_{c2}$ we
obtain the relations:
\begin{eqnarray}
\langle\chi_{c_0} \vert {\cal G}(^3P_0) \vert \chi_{c0}\rangle &=&
m_c E_{\chi_{c0}}\langle\chi_{c_0} \vert {\cal O}(^3P_0) \vert \chi_{c0}\rangle
   +m_c \langle\chi_{c_0} \vert {\cal F}(^3P_0) \vert \chi_{c0}\rangle,
   \nonumber\\
\langle\chi_{c_2} \vert {\cal G}(^3P_2) \vert \chi_{c2}\rangle &=&
m_c E_{\chi_{c2}}\langle\chi_{c_2} \vert {\cal O}(^3P_2) \vert \chi_{c2}\rangle
   +m_c \langle\chi_{c_2} \vert {\cal F}(^3P_2) \vert \chi_{c2}\rangle,
\end{eqnarray}
and the binding energy is defined as:
\begin{equation}
E_{\chi_{c0}}=m_{\chi_{c0}} - 2m_c, \ \ \
E_{\chi_{c2}}=m_{\chi_{c2}}-2m_c.
\end{equation}
It should be noted that the difference $E_{\chi_{c2}}-E_{\chi_{c0}}$ is at order higher than
$v^2$, hence the above relations are in consistent with those in Eq.(27). To study the numerical
impact we define the following parameters:
\begin{eqnarray}
a_8 &=& \frac{ \langle\chi_{c0} \vert {\cal F}(^3P_0) \vert \chi_{c0}\rangle}
          {m_c \langle\chi_{c0} \vert {\cal O}(^3P_0) \vert \chi_{c0}\rangle} =
           \frac{ \langle\chi_{c2} \vert {\cal F}(^3P_2) \vert \chi_{c2}\rangle}
          {m_c \langle\chi_{c2} \vert {\cal O}(^3P_2) \vert \chi_{c2}\rangle}\cdot( 1
            +{\cal O}(v^2)), \nonumber\\
a_F &=&\frac{ \langle\chi_{c2} \vert {\cal G}'(^3P_2) \vert \chi_{c2}\rangle}
          {{m_c}^2 \langle\chi_{c2} \vert {\cal O}(^3P_2) \vert \chi_{c2}\rangle}.
\end{eqnarray}
These parameters are order of $v^2$. With these parameters and with the relations in Eq.(31)
the decay widths can be written:
\begin{eqnarray}
\Gamma (\chi_{c0} \longrightarrow \gamma \gamma ) &=& \frac{6\pi Q_c^4\alpha^2}{{m_c}^4} \langle \chi_{c0} \vert {\cal O} (^3P_0) \vert \chi_{c0} \rangle
 \cdot \left\{ 1+0.1787\frac{\alpha_s(m_c)}{\pi} -2.333\frac{E_{\chi_{c0}}}{m_c}
  -3a_8\right\} +{\cal O}(v^6), \nonumber\\
\Gamma (\chi_{c2} \longrightarrow \gamma \gamma ) &=& \frac{8\pi Q_c^4\alpha^2}{5{m_c}^4} \langle \chi_{c2} \vert {\cal O} (^3P_2) \vert \chi_{c2} \rangle
\cdot\left\{ 1-5.333\frac{\alpha_s(m_c)}{\pi}-1.024\frac{E_{\chi_{c2}}}{m_c}
  -2.2738 a_8 -1.6905 a_F \right\} \nonumber\\
&& \hspace{3cm} + {\cal O}(v^6),
\end{eqnarray}
where we have added the one-loop correction to the contribution
at leading order of $v$\cite{Barb}. If we take $m_c=1.5$GeV and $\alpha_s(m_c)\approx 0.3$,
we obtain by using experimental values of hadron masses:
\begin{eqnarray}
\Gamma (\chi_{c0} \longrightarrow \gamma \gamma ) & \approx & \frac{6\pi Q_c^4\alpha^2}{{m_c}^4} \langle \chi_{c0} \vert {\cal O} (^3P_0) \vert \chi_{c0} \rangle
 \cdot \left\{ 1+0.017-0.645-0.3\right\} +{\cal O}(v^6), \nonumber\\
\Gamma (\chi_{c2} \longrightarrow \gamma \gamma ) & \approx & \frac{8\pi Q_c^4\alpha^2}{5{m_c}^4} \langle \chi_{c2} \vert {\cal O} (^3P_2) \vert \chi_{c2} \rangle
\cdot\left\{ 1-0.510 -0.380 -0.227-0.169 \right\} + {\cal O}(v^6),
\end{eqnarray}
where we have taken $a_8=a_F=0.1$. In the above equations, the numbers in brackets
correspond to the terms in brackets in Eq.(34), respectively. From these numbers,
we see that the effect of binding energies is very significant, also the effects
from color octet contributions and from mixing with the F-wave state are not small,
although their effects can not be determined precisely. However, our results
show that each correction at the next-to-leading order is significant. Unless
among those corrections some cancellation happens, they can not be neglected.
\par

\par\vskip20pt
\noindent
{\bf 4. Summary and conclusion}
\par\vskip20pt
We have presented a systematic study of high order corrections in
two photon decays of $\chi_{c0,2}$. Using NRQCD velocity power counting rule,
implicit expansion of decay width in relative velocity $v$ is achieved. We calculated not only first
order relativistic corrections but also corrections from suppressed component
$\vert (c \bar c)_8 g \rangle$ in which $c \bar c$ pair is in a color octet
state. The effect of color octet has been studied extensively in inclusive
processes. In this work we have studied the effect in exclusive processes.
Unlike most previous theoretical calculations in which
nonrelativistic wave function is employed, our calculation is based on a
field theoretic framework of NRQCD. The results are expressed by products of
perturbatively calculable short distance coefficients and nonperturbative
matrix elements which concern with initial bound states. In the case of
$\chi_{c2}$, a correction from orbital angular momentum $l=3$
is also obtained in our calculation. In previous calculations only relativistic
corrections can be estimated.
\par
Unfortunately, the matrix elements at the next-to-leading order are not known at all,
although some relations among them can be obtained  by symmetries and
by using equation of motion.  This prevents us from numerical predictions for
the decays. However, numerical impact of these corrections can be estimated
at certain level and with this estimation the effect of the corrections
is significant and can be not neglected. It would be interesting to study
these matrix elements by nonperturbative methods, especially by lattice QCD.
With a nonperturbative method one may determine how important the color octet
component is.

\par\vfil\eject

\end{document}